\documentclass{article}%
\usepackage{aip}
\usepackage{amsmath}
\usepackage{graphicx}
\usepackage{amsfonts}
\usepackage{amssymb}%
\providecommand{\U}[1]{\protect\rule{.1in}{.1in}}
\providecommand{\U}[1]{\protect\rule{.1in}{.1in}}
\providecommand{\U}[1]{\protect\rule{.1in}{.1in}}
\providecommand{\U}[1]{\protect\rule{.1in}{.1in}}
\providecommand{\U}[1]{\protect\rule{.1in}{.1in}}
\providecommand{\U}[1]{\protect\rule{.1in}{.1in}}
\providecommand{\U}[1]{\protect\rule{.1in}{.1in}}
\providecommand{\U}[1]{\protect\rule{.1in}{.1in}}

\begin{document}

\title{Definition of the Riesz Derivative and its Application to Space Fractional
Quantum Mechanics}
\author{Sel\c{c}uk \c{S}. Bayin\\Middle East Technical University\\Institute of Applied Mathematics\\Ankara TURKEY 06800}
\date{\today}
\maketitle

\begin{abstract}
We investigate and compare different representations of the Riesz derivative,
which plays an important role in anomalous diffusion and space fractional
quantum mechanics. In particular, we show that a certain representation of the
Riesz derivative, $R_{x}^{\alpha},$ that is generally given as also valid for
$\alpha=1$, behaves no differently than the other definition given in terms of
its Fourier transform. In the light of this, we discuss the $\alpha
\rightarrow1$ limit of the space fractional quantum mechanics and its consistency.


\end{abstract}

\section{ Introduction}

Fractional calculus is an effective tool to study nonlocal and memory effects
in physics. Successful applications to anomalous diffusion and evolution
problems [1--4] were immediately followed by applications to quantum mechanics
[5--12]. \ In particular, Laskin's space fractional quantum mechanics is
intriguing since it also follows from Feynman's path integral approach over
L\'{e}vy paths [5]. One of the first solutions of the space fractional
Schr\"{o}dinger equation was given for the infinite square well problem [6].
Despite its simplicity, this solution is important since it is basically the
prototype of a quantum detector with internal degrees of freedom. Recently,
this solution has also been the subject of some \textit{controversy} and was
used to point to the potential existence of an \textit{inconsistency} in the
solutions obtained by the piecewise method [10--13, 15, 16]. The proposed
inconsistency argument was based on the evaluation of a certain integral,
which resulted when the solution for the box problem obtained by the piecewise
method was substituted back into the Schr\"{o}dinger equation [15, 16]. One of
the crucial elements of the space fractional quantum mechanics is the Riesz
derivative. We have shown that a particular representation of the Riesz
derivative that accommodates analytic continuation can be used to evaluate the
integral in question, thus resolving the so-called \textit{inconsistency
problem }[10--13]. Misleading conclusions regarding the Laskin's space
fractional quantum mechanics [5] often results when one ignores the basic
assumptions and restrictions involved in the use of the Riesz derivative [15--21].

A crucial part of the space fractional quantum mechanics is the Riesz
derivative operator, $R_{x}^{\alpha}$, which satisfies the fractional
diffusion equation:
\begin{equation}
\frac{\partial p_{L}(x,t;\alpha)}{\partial t}-\sigma_{\alpha}R_{x}^{\alpha
}p_{L}(x,t;\alpha)=0,
\end{equation}
where $P_{L}(x,t)$ is the $\alpha-$stable L\'{e}vy distribution and $\alpha,$
$0<\alpha\leq2,$ is called the L\'{e}vy index. The $\alpha-$ stable L\'{e}vy
distribution with $0<\alpha<2$ has finite moments of order $\mu<\alpha$ and
infinite moments for higher orders [5]. The Gaussian distribution, $\alpha=2$,
is also stable with moments of all orders. In space fractional quantum
mechanics, the existence of average position, $\left\langle x\right\rangle ,$
and momentum, $\left\langle p\right\rangle $, of a particle demands that the
moments of first order exist [5]. In this regard,$\ \alpha$ has to be
restricted to the range $1<\alpha\leq2$.

Even though the problems regarding a particular integral in the infinite box
solution, which was the basis of the inconsistency arguments, has been
resolved for the range $1<\alpha\leq2,$ potential issues regarding the
$\alpha\rightarrow1$ limit of the solutions and its connection with the
particular solution of the space fractional Schr\"{o}dinger equation for
$\alpha=1$ need to be clarified [20]. As the upper bound, $\alpha=2,$ is
approached, the space fractional Schr\"{o}dinger equation approaches smoothly
to the ordinary Schr\"{o}dinger equation for the classical particle. However,
we can not say the same thing as $\alpha$ approaches the lower bound. We will
discuss in the last section that within the context of\ the Schr\"{o}dinger
theory, the interpretation of the \textit{ordinary derivative} operator for
the lower bound, $\alpha=1$, and the corresponding Hamiltonian is dubious [17, 18].

In literature, arguments about the $\alpha=1$ case are usually carried over
another representation of the Riesz derivative, which is usually given as also
valid for $\alpha=1$ [3, 21--26]. To investigate this in detail, in Section II
we start with a brief review of the Riesz derivative, which is generally given
in terms of its Fourier transform and valid for the range $0<\alpha\leq$ $2,$
$\alpha\neq1.$ In Section III, we continue with another representation of the
Riesz derivative, which is generally given in literature as valid over the
entire range, $0<\alpha\leq$ $2,$ including $\alpha=1$ [3, 21--26]. We
scrutinize its derivation and its Fourier transform, and on the contrary to
common opinion, we show that its behavior at and near $\alpha=1$ is no
different than the previous definition. Finally, in Section IV we have
conclusions and discuss the implications of our results in terms of Laskin's
space fractional quantum mechanics. We argue that the nonlocality implied by
the Riesz derivative is of different nature than the nonlocatily of the
particular solution of the Schr\"{o}dinger equation for $\alpha=1.$

\section{Definition of the Riesz derivative}

Riesz derivative, $R_{x}^{\alpha}f(x),$ is usually defined in terms of its
Fourier transform as%
\begin{equation}
\mathcal{F}\left\{  R_{x}^{\alpha}f(x)\right\}  =-\left\vert \omega\right\vert
^{\alpha}F(\omega),\text{ }\alpha>0.
\end{equation}
The fact that $\left\vert \omega\right\vert ^{\alpha}$ is not an analytic
function does not allow one to use complex contour integral theorems. In space
fractional quantum mechanics, one usually encounters real singular integrals
like
\begin{equation}
I=\int_{-\infty}^{\infty}\left\vert \omega\right\vert ^{\pm\alpha}%
f(\omega)d\omega,
\end{equation}
where $f(\omega)$ is a complex valued even function with finite number of
singular points on the real axis. One can also write $I$ as
\begin{equation}
I=2\int_{0}^{\infty}\left\vert \omega\right\vert ^{\pm\alpha}f(\omega)d\omega.
\end{equation}
However, a common source of error is in dropping the absolute value sign and
then evaluating the integral :\ \ \
\begin{equation}
I^{\prime}=2\int_{0}^{\infty}\omega^{\pm\alpha}f(\omega)d\omega,
\end{equation}
via the complex contour integral theorems [27]. The last step naturally alters
the analytic structure of the Riesz derivative and as in the box problem leads
to misleading results [15, 16, 28]. In such situations, using the original
expression of the Riesz derivative [10--13] that accommodates analytic
continuation allows correct implementation of the contour integral theorems,
thus resolving the controversy.

Using the Fourier transforms [13]:
\begin{align}
\mathcal{F}\left\{  _{-\infty}D_{x}^{\alpha}f(x)\right\}   &  =(i\omega
)^{\alpha}F(\omega),\text{ }\alpha>0\\
\mathcal{F}\left\{  _{\infty}D_{x}^{\alpha}f(x)\right\}   &  =(-i\omega
)^{\alpha}F(\omega),\text{ }\alpha>0,
\end{align}
where $F(\omega)=\mathcal{F}\left\{  f(x)\right\}  ,$ we define the derivative%
\begin{equation}
D_{x}^{\alpha}f(x)=(_{-\infty}D_{x}^{\alpha}+_{\infty}D_{x}^{\alpha})f(x),
\end{equation}
the Fourier transform of which is%
\begin{equation}
\mathcal{F}\left\{  D_{x}^{\alpha}f(x)\right\}  =((i\omega)^{\alpha}%
+(-i\omega)^{\alpha})F(\omega),\text{ }\alpha>0.
\end{equation}
For real $\omega$, the Fourier transform, $\mathcal{F}\left\{  D_{x}^{\alpha
}f(x)\right\}  ,$ can be written as%
\begin{equation}
\mathcal{F}\left\{  D_{x}^{\alpha}f(x)\right\}  =\left\vert \omega\right\vert
^{\alpha}2\cos(\alpha\pi/2)F(\omega),\text{ }\alpha>0.
\end{equation}
From here, it is seen that $D_{x}^{\alpha}f(x)$ does not have the desired
Fourier transform for neither $\alpha=1$ nor $\alpha=2$, that is,
\begin{align}
\mathcal{F}\left\{  D_{x}^{1}f(x)\right\}   &  \neq i\omega F(\omega),\text{
}\\
\mathcal{F}\left\{  D_{x}^{2}f(x)\right\}   &  \neq(i\omega)^{2}%
F(\omega)=-\left\vert \omega\right\vert ^{2}F(\omega).
\end{align}
In this regard, the Riesz derivative is defined with a minus sign as [13,
24--26]
\begin{equation}
R_{x}^{\alpha}f(x)=-\frac{(_{-\infty}D_{x}^{\alpha}+_{\infty}D_{x}^{\alpha
})f(x)}{2\cos(\alpha\pi/2)},\label{10}%
\end{equation}
where its Fourier transform becomes%
\begin{equation}
\mathcal{F}\left\{  R_{x}^{\alpha}f(x)\right\}  =-\frac{(i\omega)^{\alpha
}+(-i\omega)^{\alpha}}{2\cos(\alpha\pi/2)}F(\omega).\label{10a}%
\end{equation}
This form of the Riesz derivative allows analytic continuation and thus the
correct implementation of the complex contour integral theorems becomes
possible [10--13, 27]. For real $\omega,$ $\mathcal{F}\left\{  R_{x}^{\alpha
}f(x)\right\}  $ [Eq. (\ref{10a})] can be written as
\begin{equation}
\mathcal{F}\left\{  R_{x}^{\alpha}f(x)\right\}  =-\left\vert \omega\right\vert
^{\alpha}F(\omega).\label{10b}%
\end{equation}
This definition of the Riesz derivative has the desired Fourier transform for
$\alpha=2$, but it still does not reproduce the standard result for
$\alpha=1.$ Therefore, the above definition is generally written as valid for
$0<\alpha\leq2,$ $\alpha\neq1.$

In space fractional quantum mechanics, the\ $\alpha=2$ case corresponds to the
Schr\"{o}dinger equation for a massive nonrelativistic particle, while the
$\alpha=1$ case needs to be scrutinized carefully both on physical and
mathematical grounds. In the following section, we investigate another
representation of the Riesz derivative that is given in literature as also
valid for $\alpha=1$ and thus written as good for the full range $0<\alpha
\leq2$ [3, 21--26].

\section{Another Representation of the Riesz Derivative}

We start with the formula [1--3, 27]%
\begin{equation}
_{-\infty}D_{x}^{\alpha}f(x)=\frac{d^{2}}{dx^{2}}\left[  _{-\infty}%
I_{x}^{2-\alpha}f(x)\right]  ,\text{ }1<\alpha<2,
\end{equation}
and write%
\begin{align}
_{-\infty}D_{x}^{\alpha}f(x)  &  =_{-\infty}I_{x}^{-\alpha}f(x)\\
&  =\frac{1}{\Gamma(2-\alpha)}\frac{d^{2}}{dx^{2}}\int_{-\infty}^{x}%
\frac{f(x^{\prime})}{(x-x^{\prime})^{\alpha-1}}dx^{\prime}\\
&  =\frac{1}{\Gamma(2-\alpha)}\frac{d^{2}}{dx^{2}}\int_{0}^{\infty}%
\xi^{-\alpha+1}f(x-\xi)d\xi.
\end{align}
Using the relations:%
\begin{align}
\xi^{-\alpha+1}  &  =(\alpha-1)\int_{\xi}^{\infty}\frac{d\eta}{\eta^{\alpha}%
},\\
\frac{\partial^{2}f(x-\xi)}{\partial x^{2}}  &  =\frac{\partial^{2}f(x-\xi
)}{\partial\xi^{2}},
\end{align}
we can write%
\begin{equation}
_{-\infty}D_{x}^{\alpha}f(x)=\frac{(\alpha-1)}{\Gamma(2-\alpha)}\int
_{0}^{\infty}\frac{\partial^{2}f(x-\xi)}{\partial\xi^{2}}\left[  \int_{\xi
}^{\infty}\frac{d\eta}{\eta^{\alpha}}\right]  d\xi,\text{ }1<\alpha<2.
\end{equation}
Integrating by parts twice yields%
\begin{equation}
_{-\infty}D_{x}^{\alpha}f(x)=-\frac{\ \alpha}{\Gamma(1-\alpha)}\int
_{0}^{\infty}\frac{f(x-\xi)-f(x)}{\xi^{\alpha+1}}d\xi,\text{ }1<\alpha<2.
\end{equation}
Following similar steps, we obtain%
\begin{equation}
_{\infty}D_{x}^{\alpha}f(x)=-\frac{\ \alpha}{\Gamma(1-\alpha)}\int_{0}%
^{\infty}\frac{f(x+\xi)-f(x)}{\xi^{\alpha+1}}d\xi,\text{ }1<\alpha<2.
\end{equation}
Combining these in Equation (\ref{10}) we obtain another representation of the
Riesz derivative:%
\begin{equation}
R_{x}^{\alpha}f(x)=\frac{\Gamma(1+\alpha)\sin\alpha\pi/2}{\pi}\int_{0}%
^{\infty}\frac{f(x+\xi)-2f(x)+f(x-\xi)}{\xi^{1+\alpha}}d\xi,\text{ }%
1<\alpha<2. \label{22}%
\end{equation}

Using similar steps, one can show that an identical relation results for the
range $0<\alpha<1$ [25]. In Literature, this representation is commonly used
as another representation of the Riesz derivative that is also regular at
$\alpha=1,$ thus written as good for the entire range $0<\alpha\leq2$ [3, 21
--26]. Since this point has been a source of major misunderstanding and misuse
of the Riesz derivative, we will analyze the end points\ carefully and hope to
clear any misconceptions that exists.

\subsection{Riesz Derivative for $0<\alpha<1$ and$\ 1<\alpha<2$}

Before we discuss the behavior of the Riesz derivative at the end points,
$\alpha=1$ and $\alpha=2,$ we concentrate on the derivative, $\widetilde
{D}_{x}^{\alpha}f(x):$
\begin{align}
\widetilde{D}_{x}^{\alpha}f(x)  &  =\frac{(_{-\infty}D_{x}^{\alpha}+_{\infty
}D_{x}^{\alpha})f(x)}{2\cos(\alpha\pi/2)}\label{24}\\
&  =-\frac{\Gamma(1+\alpha)\sin\alpha\pi/2}{\pi}\int_{0}^{\infty}\frac
{f(x+\xi)-2f(x)+f(x-\xi)}{\xi^{1+\alpha}}d\xi,\text{ }%
\end{align}
which has been obtained for the ranges $0<\alpha<1$ and $1<\alpha<2$,
separately. Note that we have taken out the minus sign in front of Equation
(\ref{10}) and Equation (\ref{22}) which was inserted by hand in the first
place. In other words, $-\widetilde{D}_{x}^{\alpha}f(x)$ is the second
representation of the Riesz derivative (\ref{22}):
\begin{equation}
R_{x}^{\alpha}f(x)=-\widetilde{D}_{x}^{\alpha}f(x). \label{25}%
\end{equation}
Evaluating the Fourier transform, $\mathcal{F}\left\{  \widetilde{D}%
_{x}^{\alpha}f(x)\right\}  ,$ with respect to $x,$ and using the relations:%
\begin{align}
\frac{\alpha}{\Gamma(1-\alpha)}  &  =-\frac{1}{\Gamma(-\alpha)}=\Gamma
(1+\alpha)\frac{\sin\alpha\pi}{\pi},\\
\text{ }\mathcal{F}\left\{  f(x-\xi)\right\}   &  =e^{-i\omega\xi}F(\omega),
\end{align}
we can write%
\begin{align}
&  \mathcal{F}\left\{  \widetilde{D}_{x}^{\alpha}f(x)\right\} \\
&  =-\frac{\ \alpha}{\Gamma(1-\alpha)}\left[  \mathcal{F}\left\{  \int
_{0}^{\infty}\frac{f(x-\xi)}{\xi^{1+\alpha}}d\xi\right\}  +\mathcal{F}\left\{
\int_{0}^{\infty}\frac{f(x+\xi)}{\xi^{1+\alpha}}d\xi\right\}  -2\mathcal{F}%
\left\{  \int_{0}^{\infty}\frac{f(x)}{\xi^{1+\alpha}}d\xi\right\}  \right]
,\\
&  =-\frac{\ \alpha}{\Gamma(1-\alpha)}\left[  \int_{0}^{\infty}\frac
{\mathcal{F}\left\{  f(x-\xi)\right\}  }{\xi^{1+\alpha}}d\xi+\int_{0}^{\infty
}\frac{\mathcal{F}\left\{  f(x+\xi)\right\}  }{\xi^{1+\alpha}}d\xi-2\int
_{0}^{\infty}\frac{\mathcal{F}\left\{  f(x)\right\}  }{\xi^{1+\alpha}}%
d\xi\right] \\
&  =-\frac{\ \alpha\ F(\omega)\ }{\Gamma(1-\alpha)}\left[  \int_{0}^{\infty
}\frac{e^{-i\omega\xi}}{\xi^{1+\alpha}}d\xi+\int_{0}^{\infty}\frac
{e^{i\omega\xi}}{\xi^{1+\alpha}}d\xi-2\int_{0}^{\infty}\frac{1}{\xi^{1+\alpha
}}d\xi\right]  ,\label{29}\\
&  =-\frac{\ \alpha\ F(\omega)\ }{\Gamma(1-\alpha)}\left[  I_{1}+I_{2}%
-I_{3}\right]  ,\text{ }1<\alpha<2,\text{ }0<\alpha<1, \label{29b}%
\end{align}
where
\begin{equation}
I_{1}=\int_{0}^{\infty}\frac{e^{-i\omega\xi}}{\xi^{1+\alpha}}d\xi,\text{
}I_{2}=\int_{0}^{\infty}\frac{e^{i\omega\xi}}{\xi^{1+\alpha}}d\xi,\text{
}I_{3}=2\int_{0}^{\infty}\frac{1}{\xi^{1+\alpha}}d\xi. \label{29c}%
\end{equation}
We have assumed that the integral over $\xi$ and the integral from the Fourier
transform with respect to $x$ can be interchanged. The three integrals [Eq.
(\ref{29c})]: are singular and do not exist in the Riemann sense. However,
using analytic continuation and along with an appropriate contour, we can
regularize these divergent integrals and show that a meaningful result for
their sum exists.%
\begin{figure}
[ptb]
\begin{center}
\includegraphics[
height=1.9709in,
width=2.5694in
]%
{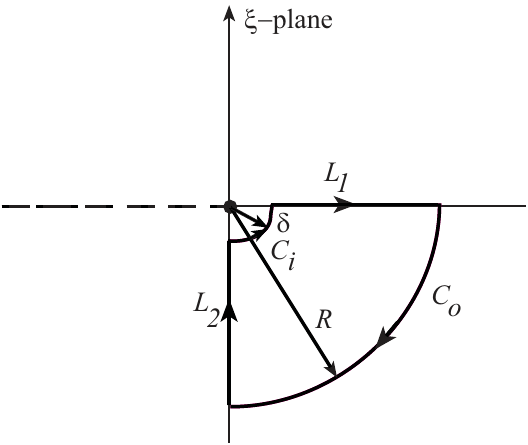}%
\caption{Contour for $I_{1}.$ The dashed line indicates the branch cut and the
pole is at the center.}%
\end{center}
\end{figure}

Starting with $I_{1},$ we use the contour in Fig. 1 and evaluate the contour
integral%
\begin{equation}%
{\displaystyle\oint_{C}}
\frac{e^{-i\omega\xi}}{\xi^{1+\alpha}}d\xi,
\end{equation}
where $\xi$ is now in the complex $\xi-$plane and the contour $C$ has the
parts $C_{i},$ $C_{0},$ $L_{1}$ and $L_{2}.$ Since there are no singularities
inside $C,$ we can write [27]%
\begin{align}%
{\displaystyle\oint_{C}}
\frac{e^{-i\omega\xi}}{\xi^{1+\alpha}}d\xi &  =0,\label{33a}\\%
{\displaystyle\oint_{C_{0}}}
\frac{e^{-i\omega\xi}}{\xi^{1+\alpha}}d\xi+%
{\displaystyle\oint_{C_{i}}}
\frac{e^{-i\omega\xi}}{\xi^{1+\alpha}}d\xi+%
{\displaystyle\oint_{L_{1}}}
\frac{e^{-i\omega\xi}}{\xi^{1+\alpha}}d\xi+%
{\displaystyle\oint_{L_{2}}}
\frac{e^{-i\omega\xi}}{\xi^{1+\alpha}}d\xi &  =0, \label{33b}%
\end{align}
where the integral over $C_{0}$ is to be evaluated in the limit $R\rightarrow
\infty$ and the integral $C_{i}$ is to be evaluated in the limit
$\delta\rightarrow0.$ In these limits, the integral over $L_{1}:$%
\[%
{\displaystyle\oint_{L_{1}}}
\frac{e^{-i\omega\xi}}{\xi^{1+\alpha}}d\xi,
\]
is the needed integral, $I_{1},$ while the integral over $L_{2}$ can be
evaluated as%
\begin{equation}%
{\displaystyle\oint_{L_{2}}}
\frac{e^{-i\omega\xi}}{\xi^{1+\alpha}}d\xi=-(i\omega)^{\alpha}\Gamma(-\alpha).
\label{34}%
\end{equation}
For the integral over $C_{0}$ we can put an upper bound as%
\begin{align}
I_{C_{0}}  &  =%
{\displaystyle\oint_{C_{0}}}
\frac{e^{-i\omega\xi}}{\xi^{1+\alpha}}d\xi\\
&  =\int_{0}^{-\pi/2}i\frac{e^{-i\alpha\theta}}{R^{\alpha}}e^{-i\omega
R\cos\theta+\omega R\sin\theta}d\theta\\
&  \leq\int_{0}^{-\pi/2}\left\vert i\frac{e^{-i\alpha\theta}}{R^{\alpha}%
}e^{-i\omega R\cos\theta+\omega R\sin\theta}\right\vert d\theta\\
&  \leq\frac{1}{R^{\alpha}}\int_{0}^{-\pi/2}e^{\omega R\sin\theta}d\theta
\leq\frac{1}{R^{\alpha}}\int_{0}^{-\pi/2}e^{\omega R(2\theta/\pi)}%
d\theta\label{35}\\
&  \leq\frac{1}{R^{\alpha}}\frac{\pi}{2\omega R}\left[  e^{(2\omega
R/\pi)\theta}\right]  _{0}^{-\pi/2}\leq\frac{1}{R^{\alpha+1}}\frac{\pi
}{2\omega}\left[  e^{-\omega R}-1\right]  ,
\end{align}
where in Equation (\ref{35}) we have used the the inequality [27]
\[
2\theta/\pi\geq\sin\theta,\text{ }\theta\in\left[  -\pi/2,0\right]  .
\]
Thus, in the limit as $R\rightarrow\infty,$ the integral $\left\vert I_{C_{0}%
}\right\vert $ goes to $0$. On the other hand, the integral over $C_{i}:$%
\begin{align}
I_{C_{i}}  &  =%
{\displaystyle\oint_{C_{i}}}
\frac{e^{-i\omega\xi}}{\xi^{1+\alpha}}d\xi\\
&  =\int_{-\pi/2}^{0}i\delta^{-\alpha}e^{-i\alpha\theta}e^{-i\omega\delta
\cos\theta+\omega\delta\sin\theta}d\theta\\
&  =\frac{i}{\delta^{\alpha}}\int_{-\pi/2}^{0}e^{-i\alpha\theta}\left[
e^{\omega\delta(\sin\theta-i\cos\theta})\right]  d\theta,
\end{align}
diverges as $\delta\rightarrow0.$ Expanding the exponential inside the square
brackets and integrating term by term we can write $I_{C_{i}}$ explicitly in
terms of $\delta$ as%
\begin{equation}
I_{C_{i}}=\frac{1}{\delta^{\alpha}\alpha}\left[  (e^{i\alpha\pi/2}%
-1)-i(\omega\delta)\frac{\sin(\alpha-1)\pi/2}{(\alpha-1)}+O(\omega^{2}%
\delta^{2})\right]  .
\end{equation}
The divergent part of $I_{C_{i}}$ is now explicitly written in terms of
$\delta$ as%
\begin{equation}
\lim_{\delta\rightarrow0}I_{C_{i}}=-\frac{1}{\delta^{\alpha}\alpha
}(1-e^{i\alpha\pi/2}). \label{42}%
\end{equation}
Combining these results in Equation (\ref{33b}), we obtain $I_{1}$ as%
\begin{equation}
I_{1}=(i\omega)^{\alpha}\Gamma(-\alpha)+\frac{1}{\delta^{\alpha}\alpha
}(1-e^{i\alpha\pi/2}). \label{43}%
\end{equation}%
\begin{figure}
[ptb]
\begin{center}
\includegraphics[
height=1.8555in,
width=2.0722in
]%
{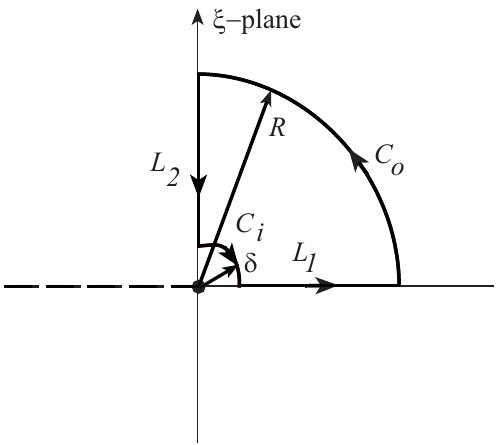}%
\caption{Contour for $I_{2}.$}%
\end{center}
\end{figure}
Similarly, but with the contour in Fig. 2, and in the limit $\delta
\rightarrow0,$ we obtain $I_{2}$ as
\begin{equation}
I_{2}=(-i\omega)^{\alpha}\Gamma(-\alpha)+\frac{1}{\delta^{\alpha}\alpha
}(1-e^{-i\alpha\pi/2}). \label{44}%
\end{equation}
Since the results for $I_{1}$ and $I_{2}$ are valid for both $1<\alpha<2$ and
$0<\alpha<1,$ separately, we can deduce the value of $I_{3}$ from $I_{1}$ and
$I_{2}$ by setting $\omega$ to zero. However, we first we add the two graphs
(Fig. 3) to write%
\begin{equation}
I_{L_{1}+L_{2}+C_{0}+C_{i}}+I_{L_{1}^{\prime}+L_{2}^{\prime}+C_{0}^{\prime
}+C_{i}^{\prime}}=0.
\end{equation}
When $\omega=0,$ the integrals over $L_{1}$ and $L_{1}^{\prime}$ are equal and
thus give the needed integral $I_{3}$ as
\begin{align}
I_{3}  &  =\left[
{\displaystyle\oint_{L_{1}}}
\frac{e^{-i\omega\xi}}{\xi^{1+\alpha}}d\xi+%
{\displaystyle\oint_{L_{1}^{\prime}}}
\frac{e^{i\omega\xi}}{\xi^{1+\alpha}}d\xi\right]  _{\omega=0}\\
&  =2\int_{0}^{\infty}\frac{1}{\xi^{1+\alpha}}d\xi.
\end{align}
Therefore,
\begin{equation}
I_{3}=-\left[
{\displaystyle\oint_{L_{2}}}
\frac{e^{-i\omega\xi}}{\xi^{1+\alpha}}d\xi+%
{\displaystyle\oint_{L_{2}^{\prime}}}
\frac{e^{i\omega\xi}}{\xi^{1+\alpha}}d\xi\right]  -\left[
{\displaystyle\oint_{C_{i}^{\prime}}}
\frac{e^{-i\omega\xi}}{\xi^{1+\alpha}}d\xi+%
{\displaystyle\oint_{C_{i}}}
\frac{e^{i\omega\xi}}{\xi^{1+\alpha}}d\xi\right]  .
\end{equation}
Using Equations (\ref{34}, \ref{42}--\ref{44}) we write%
\begin{equation}
I_{3}=\left[  (i\omega)^{\alpha}+(-i\omega)^{\alpha}\right]  \Gamma
(-\alpha)+\left[  \frac{1}{\delta^{\alpha}\alpha}(1-e^{i\alpha\pi/2})+\frac
{1}{\delta^{\alpha}\alpha}(1-e^{-i\alpha\pi/2})\right]  +0(\omega^{2}%
\delta^{2}).
\end{equation}
Setting $\omega=0$ and taking the limits $R\rightarrow\infty$ and
$\delta\rightarrow0,$ we finally obtain $I_{3}$ as%
\begin{equation}
I_{3}=\frac{1}{\delta^{\alpha}\alpha}(2-e^{i\alpha\pi/2}-e^{-i\alpha\pi/2}).
\end{equation}%
\begin{figure}
[ptb]
\begin{center}
\includegraphics[
height=2.1162in,
width=2.0182in
]%
{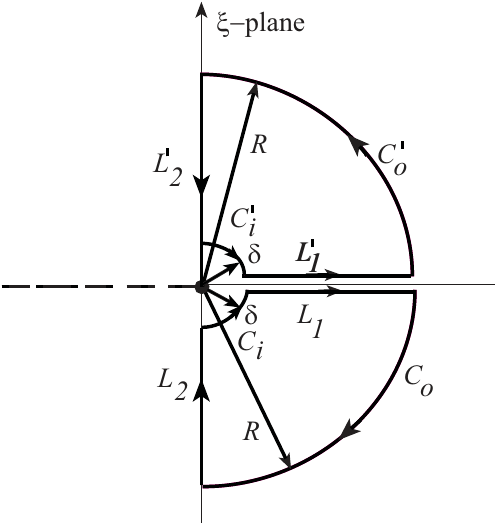}%
\caption{Combined paths for $I_{3}$.}%
\end{center}
\end{figure}
Combining these in Equation (\ref{29}) and using the relation
\begin{align}
\Gamma(-\alpha)\Gamma(1+\alpha)  &  =-\frac{\pi}{\sin\pi\alpha}\\
&  =-\frac{\pi}{2\sin\pi\alpha/2\cos\pi\alpha/2},
\end{align}
we finally obtain%
\begin{equation}
\mathcal{F}\left\{  \widetilde{D}_{x}^{\alpha}f(x)\right\}  =\frac{\left[
(i\omega)^{\alpha}+(-i\omega)^{\alpha}\right]  }{2\cos\pi\alpha/2}%
F(\omega),\text{ }0<\alpha<1,\text{ }1<\alpha<2. \label{55}%
\end{equation}
Note that the divergences as $\delta\rightarrow0$ in $I_{1}$ and $I_{2}$ are
cancelled by the divergence in $I_{3},$ thus yielding a finite result [Eq.
(\ref{55})] for the transform $\mathcal{F}\left\{  \widetilde{D}_{x}^{\alpha
}f(x)\right\}  $.$\ $For real $\omega,$ this can also be written as%
\begin{equation}
\mathcal{F}\left\{  \widetilde{D}_{x}^{\alpha}f(x)\right\}  =\left\vert
\omega\right\vert ^{\alpha}F(\omega),\text{ }0<\alpha<1,\text{ }1<\alpha<2.
\label{56}%
\end{equation}

\subsection{The $q=1$ case and the $q\rightarrow1^{\pm}$ limits}

We now investigate the $\alpha=1$ case. From Equation (\ref{55}) it is seen
that the Fourier transform $\mathcal{F}\left\{  \widetilde{D}f(x)\right\}  $
at $\alpha=1$ is $0/0,$ thus undefined. Since Equation (\ref{55}) is valid for
the ranges $0<\alpha<1$ and $1<\alpha<2,\ $separately, we can investigate the
$\alpha=1$ case as $1$ is approached from both directions. Since the
numerator, $\left[  (i\omega)^{\alpha}+(-i\omega)^{\alpha}\right]  ,$ and the
denominator, $\cos\pi\alpha/2,$ are analytic functions of $\alpha,$ we can
write their Taylor series expansions about $\alpha=1:$
\begin{align}
(i\omega)^{\alpha}  &  =i\omega\left[  1+\ln(i\omega)(\alpha-1)\right]
+0((\alpha-1)^{2}),\label{57}\\
(-i\omega)^{\alpha}  &  =-i\omega\left[  1+\ln(-i\omega)(\alpha-1)\right]
+0((\alpha-1)^{2}),\label{58}\\
\cos\pi\alpha/2  &  =-\frac{\pi}{2}(\alpha-1)+0((\alpha-1)^{2}). \label{59}%
\end{align}
For real $\omega,$ we write
\begin{equation}
i\omega=\pm i\left\vert \omega\right\vert ,\text{ }+\text{for }\omega>0\text{
and }-\text{for }\omega<0,\text{ }%
\end{equation}
and regardless of the direction of approach, we obtain%
\begin{equation}
\lim_{\alpha\rightarrow1^{\pm}}\left[  (i\omega)^{\alpha}+(-i\omega)^{\alpha
}\right]  \simeq-\left\vert \omega\right\vert \pi(\alpha-1). \label{60}%
\end{equation}
Using Equations (\ref{59}) and (\ref{60}) in Equation (\ref{55}) we finally
obtain
\begin{align}
\lim_{\alpha\rightarrow1^{\pm}}\mathcal{F}\left\{  \widetilde{D}_{x}^{\alpha
}f(x)\right\}   &  \rightarrow\frac{-\left\vert \omega\right\vert \pi
(\alpha-1)}{-2(\pi/2)(\alpha-1)}F(\omega)\\
&  =\left\vert \omega\right\vert F(\omega).
\end{align}
Therefore, the Fourier transform, $\mathcal{F}\left\{  \widetilde{D}%
_{x}^{\alpha}f(x)\right\}  ,$ can be written for the entire range,
$0<\alpha<2,$ including $\alpha=1,$ as \
\begin{equation}
\mathcal{F}\left\{  \widetilde{D}_{x}^{\alpha}f(x)\right\}  =\left\vert
\omega\right\vert ^{\alpha}F(\omega),\text{ }0<\alpha<2,\text{ }\omega\text{
\ real.}%
\end{equation}

\section{Conclusions and Space Fractional Quantum Mechanics}

Even though the Fourier transform of $\widetilde{D}_{x}^{\alpha}f(x)$ exists
in the entire range, $0<\alpha<2,$ including $\alpha=1,$ it does not reduce to
the expected ordinary derivatives neither at $\alpha=1$ nor at $\alpha=2,$
that is,
\begin{equation}
\widetilde{D}_{x}^{1}f(x)\neq\frac{df(x)}{dx},\text{ }\widetilde{D}_{x}%
^{2}f(x)\neq\frac{d^{2}f(x)}{dx^{2}}.
\end{equation}
Introducing a minus sign rectifies the situation at $\alpha=2:$%
\begin{equation}
-\widetilde{D}_{x}^{2}f(x)=\frac{d^{2}f(x)}{dx^{2}}%
\end{equation}
but the problem at $\alpha=1$ remains. In conclusion, the second
representation of the Riesz derivative [Eqs. (\ref{22} and \ref{25})] :
\begin{equation}
R_{x}^{\alpha}f(x)=-\widetilde{D}_{x}^{\alpha}f(x),
\end{equation}
has the same Fourier transform as the first definition [Eqs. (\ref{10}%
)--(\ref{10b})] in the entire range $0<\alpha\leq2:$%
\begin{equation}
\mathcal{F}\left\{  R_{x}^{\alpha}f(x)\right\}  =\mathcal{F}\left\{
-\widetilde{D}_{x}^{\alpha}f(x)\right\}  \ =-\left\vert \omega\right\vert
^{\alpha}F(\omega),\ 0<\alpha\leq2.
\end{equation}
On the contrary to the prevailing opinion, both representations exist at
$\alpha=1$ with the Fourier transform:%
\begin{equation}
\mathcal{F}\left\{  R_{x}^{\alpha=1}f(x)\right\}  =-\left\vert \omega
\right\vert F(\omega).
\end{equation}
But unlike the $\alpha=2$ case, neither representation reduces to the ordinary
derivative at $\alpha=1,$ that is,
\begin{equation}
R_{x}^{\alpha=1}f(x)\neq\frac{df(x)}{dx}.
\end{equation}

In this regard, the Riesz derivative is usually defined as [25]
\begin{align}
R_{x}^{\alpha}f(x)  &  =-\frac{(_{-\infty}D_{x}^{\alpha}+_{\infty}%
D_{x}^{\alpha})f(x)}{2\cos(\alpha\pi/2)},\text{ }0<\alpha\leq2,\text{ }%
\alpha\neq1,\label{70a}\\
R_{x}^{\alpha}f(x)  &  =\frac{df(x)}{dx}\text{ for }\alpha=1. \label{70b}%
\end{align}
Note that for $\alpha=1,$ one can also use the integral representation [25]%
\begin{equation}
R_{x}^{\alpha=1}f(x)=\frac{d}{dx}\left[  \frac{1}{\pi}\int_{-\infty}^{\infty
}\frac{f(\xi)}{x-\xi}d\xi\right]  .
\end{equation}
In this definition [Eqs. (\ref{70a}) and (\ref{70b})] the Riesz derivative,
$R_{x}^{\alpha}f(x),$ has a discontinuity at $\alpha=1,$ which can become a
source of confusion in applications to space fractional quantum mechanics
[20]. ~In space fractional quantum mechanics, the $\alpha=2$ case corresponds
to the Schr\"{o}dinger equation for nonrelativistic particles with mass.
However, the $\alpha=1$ case deserves special attention.

First of all, the Riesz derivative, $R_{x}^{\alpha}f(x),$ in Equation
(\ref{70a}) does not reduce to the ordinary derivative $df(x)/dx$ as
$\alpha\rightarrow1.$ On the other hand, for the operator $d/dx,$ the
Schr\"{o}dinger equation:%
\begin{equation}
i\hbar\frac{\partial\Psi(x,t)}{\partial t}=H\Psi(x,t), \label{66}%
\end{equation}
and the corresponding Hamiltonian for a free particle in momentum and
configuration spaces are given, respectively, as%
\begin{align}
H  &  =pc\label{67}\\
&  =-i\hbar c\frac{d}{dx}. \label{68}%
\end{align}
In the early days of relativistic quantum mechanics, in analogy with the
classical theory, the following Hamiltonian for the relativistic free particle
was considered [29, 30]:%
\begin{equation}
H=\sqrt{p^{2}c^{2}+(m_{0}c^{2})^{2}}, \label{69}%
\end{equation}
where $m_{0}$ is the rest mass. In configuration space, the Hamiltonian
operator becomes%
\begin{equation}
H=\sqrt{-\hslash^{2}c^{2}\left(  d/dx\right)  ^{2}+m_{0}^{2}c^{4}}, \label{70}%
\end{equation}
where one immediately faces the problem of interpreting the square root
operator. Expanding the square root gives an expression that contains
derivatives of all orders, thus giving a nonlocal theory. In the classical
limit, the higher derivatives disappear thereby reducing $H$ to the well known
classical Hamiltonian operator:
\begin{equation}
H=-\frac{\hslash^{2}}{2m_{0}}\frac{d^{2}}{dx^{2}}.
\end{equation}
Such theories are\ not only very difficult to handle but also the unsymmeric
appearance of the time and the space coordinates eventually led to their
demise and opened the path to Dirac theory. In this regard, if one wants to
investigate fractional relativistic quantum mechanics, one has to start with
the Dirac theory [29, 30]. Note that it is not possible to interpret Equation
(\ref{68}) as the Hamiltonian for photons (massless particles or ultra
relativistic particles) either, since it implies a nonlocal expression for the
energy density [29, 30].

In this regard, in space fractional quantum mechanics, unlike the upper bound,
$\alpha=2$, the physical meaning of the Hamiltonian [Eq. (\ref{68})]
corresponding to $\alpha=1$ is at most dubious. Besides, even if one could
surmount the above mentioned difficulties and manage to find a solution to the
fractional Schr\"{o}dinger equation for $\alpha=1$ :%
\begin{equation}
i\hbar\frac{\partial\Psi(x,t)}{\partial t}=-i\hbar c\frac{d\Psi(x,t)}{dx},
\end{equation}
it will not be the $\alpha\rightarrow1$ limit of the solution of the Laskin's
space fractional quantum mechanics. This should not be interpreted as an
inconsistency, since the nature of the nonlocality will be different [20, 15
-- 18, 28].

The space fractional quantum mechanics via the Riesz derivative is a nonlocal
theory. However, the Riesz derivative corresponds to a particular sampling of
the function, thus it is not the only possible nonlocal theory that one could
consider. Among all possible nonlocal theories, the space fractional quantum
mechanics via the Riesz derivative has the intriguing feature that it also
follows from Feynman's path integral approach over L\'{e}vy paths.

The $\alpha-$stable L\'{e}vy distribution, $p_{L}(x,t;\alpha),$ satisfies the
fractional diffusion equation with the Riesz derivative [5]:%
\begin{equation}
\frac{\partial p_{L}(x,t;\alpha)}{\partial t}-\sigma_{\alpha}R_{x}^{\alpha
}p_{L}(x,t;\alpha)=0,
\end{equation}
\newline where $0<\alpha\leq2$ is called the L\'{e}vy index. The $\alpha-$
stable L\'{e}vy distribution with $0<\alpha<2$ has finite moments of order
$\mu<\alpha,$ but infinite moments for higher orders. The Gaussian
distribution corresponds to $\alpha=2$ and is also stable with moments of all
orders. For applications to space fractional quantum mechanics, it is
essential that the moments of first order exist. In other words, the existence
of average position and momentum of the physical particle demands that
$\alpha$ be restricted to the range $1<\alpha\leq2$ [5]. In this regard,
comparison of the $\alpha=1$ solution with the $\alpha\rightarrow1$ limit of
the Laskin's space fractional Quantum mechanics not only violates the basic
premises of the theory but also could be misinterpreted as an inconsistency.

\end{document}